\documentclass[twocolumn, 10pt, amsmath,amssymb,amsfonts]{revtex4}

\usepackage{bm,graphicx}
\usepackage{epsfig, amssymb}
\usepackage[english]{babel}

\def\be{\begin{eqnarray}}
\def\ee{\end{eqnarray}}
\def\bee{\begin{eqnarray*}}
\def\eee{\end{eqnarray*}}

\begin{document}
%

\title{Quantum algorithms for spin models and simulable gate sets for quantum computation}

\author{M. Van den Nest$^1$, W. D\"ur$^{2,3}$,  R. Raussendorf$^4$ and H. J. Briegel$^{2,3}$ }

\

\affiliation{$^1$Max-Planck-Institut f\"ur Quantenoptik, Hans-Kopfermann-Str. 1, D-85748 Garching, Germany.\\
 $^2$Institut f\"ur Quantenoptik und Quanteninformation der \"Osterreichischen Akademie der Wissenschaften, Innsbruck, Austria\\
 $^3$Institut f{\"u}r Theoretische Physik, Universit{\"a}t Innsbruck, Technikerstra{\ss}e 25, A-6020 Innsbruck
 \\ $^4$ University of British Columbia, Department of Physics and Astronomy, 6224 Agricultural Rd., Vancouver, BC V6T 1Z1, Canada}

\begin{abstract}
We present elementary mappings between classical lattice models and quantum circuits. These mappings provide a general framework to obtain efficiently simulable quantum gate sets from exactly solvable classical models. For example, we recover and generalize the simulability of Valiant's match-gates by invoking the solvability of the free-fermion eight-vertex model. Our mappings furthermore provide a systematic formalism to obtain simple quantum algorithms to approximate partition functions of lattice models in certain complex-parameter regimes. For example, we present an efficient quantum algorithm for the six-vertex model  as well as a 2D Ising-type model. We finally show that simulating our quantum algorithms on a classical computer is as hard as simulating universal quantum computation (i.e. BQP-complete).
\end{abstract}

\pacs{03.67.-a, 03.67.Lx, 75.10.Hk, 75.10.Pq, 02.70.-c}

\maketitle


{\bf 1. Introduction.---} 
Interrelations between quantum information  theory and statistical mechanics have
recently proven to be particularly fruitful. Indeed, it turns out that several
connections between these fields exist, such that techniques from either one of
these research areas may be used to gain insights in the other
\cite{Bo07,Br06,Ah07,Va08}. 

A natural context to study such connections to statistical mechanics regards the investigation of the computational power of quantum computers---in particular, understanding their potential and limits, and the borderline between classical and quantum computation. Recent findings indeed point to a fruitful cross-fertilization between this subject and statistical mechanics. For example, in \cite{Br06} it was proved that the exact solvability of the 2D Ising model can be invoked to prove that the ``toric code states'' \cite{Ki03}, regarded as resources for quantum computation, do not yield any speed-up with respect to classical computation. This indicates that, under certain circumstances, exactly solved models may be related to the efficient classical simulation of quantum computation.  In other work \cite{Ah07,Ah06}, efficient quantum algorithms are obtained to  approximate difficult quantities such as the Potts model partition function \cite {Potts}. This suggests that quantum computers may become useful tools to study problems in statistical mechanics.

Despite the existence of these interesting results, such findings typically occur as isolated mappings, and a systematic approach seems to be missing. Indeed, although recent algorithms such as \cite{Ah07} are significant achievements, they require involved arguments and are not easily generalized. One may therefore hope that quantum algorithms can be constructed to tackle statistical mechanics problems in a simpler and more systematic way. Furthermore, in statistical mechanics a lot of effort has gone in developing techniques which aim at solving spin systems, such as the Jordan-Wigner transform and the Yang-Baxter equation \cite{Ba89}. It would be highly beneficial if clearcut mappings existed which would allow one to use such established techniques to study the computational power of quantum computers in a general setting, going beyond results such as e.g. \cite{Br06}.

We believe that in the present paper significant steps are taken towards achieving
the above goals:
\begin{itemize}
\item[(i)] We present elementary, general constructions that relate arbitrary
classical lattice models with quantum circuits.
\item[(ii)] Using these mappings, we show that from any classical model where the
partition function 
can be efficiently evaluated, we obtain a class of quantum circuits that can be
efficiently simulated classically.
\item[(iii)] These mappings also immediately provide a simple framework to construct quantum algorithms that approximate the partition function of these models in certain---typically complex-valued---parameter regimes. We furthermore show that approximating the partition function of certain models is equivalent to simulating
quantum computation altogether, i.e., these models are BQP-complete \cite{noteBQP}.
\end{itemize}

More concretely, these general findings allow us to, e.g., recover and generalize a
construction of Valiant \cite{Va02}, yielding classes of simulable gate sets from
the solvability of the 8-vertex and 32-vertex model with free-fermion condition. We
furthermore obtain a class of simulable gate sets by invoking the simulability of
the 2D Ising model without external field.  We will also show that the 6-vertex
model as well as a 2D Ising-type model are, in certain complex parameter regimes,
BQP-complete models.


\

\begin{figure} {\includegraphics[width=9cm]{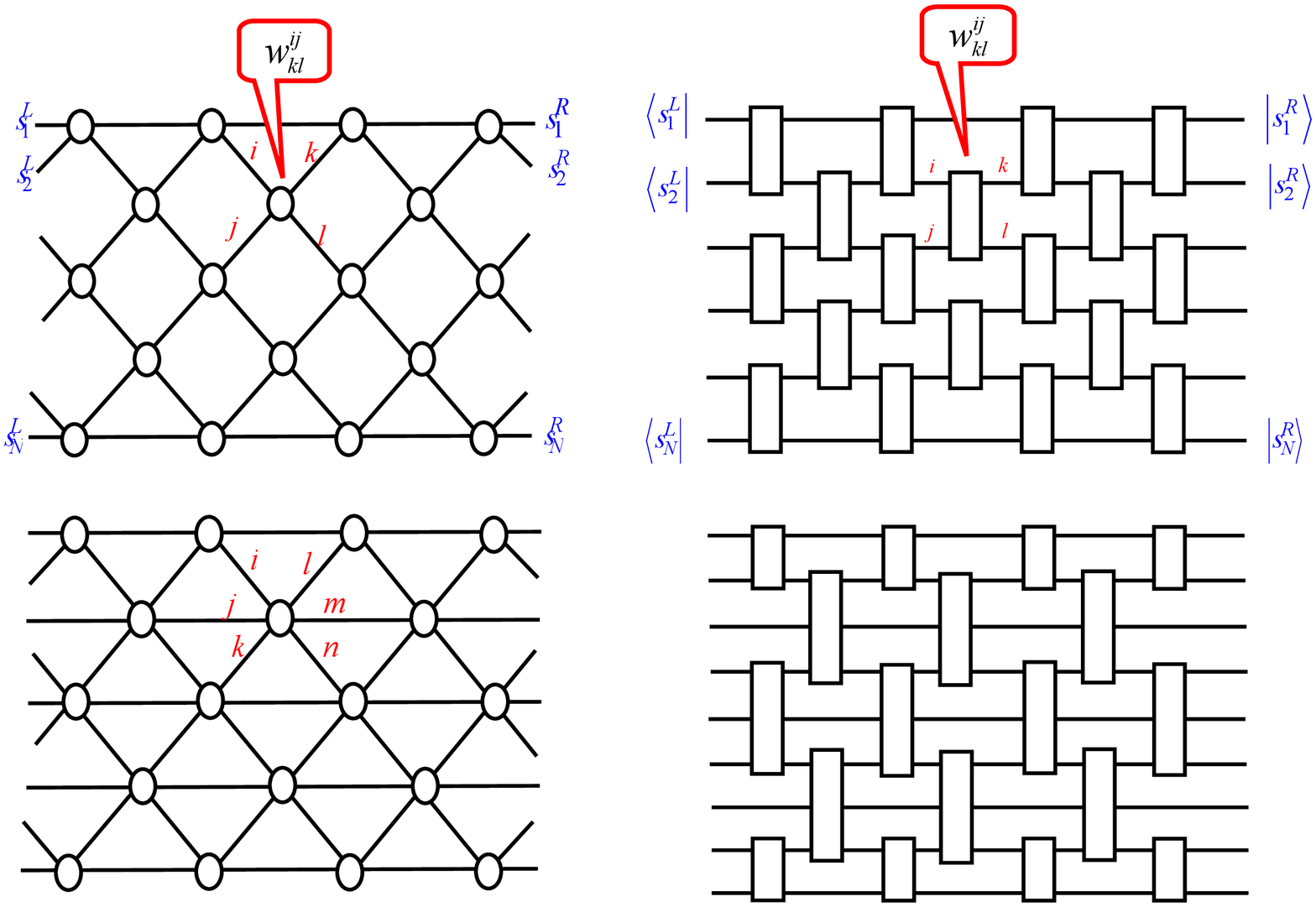}} \caption[]{\label{Fig:VM1} [Top left] Vertex model on a tilted 2D square lattice. On every edge, a $q$-state classical spin (illustrated in red) is situated. At every vertex, an interaction takes place between the four incident edges $i,j,k,l$, with corresponding Boltzmann weight $w^{ij}_{kl}$. The spins on the left and right boundary edges are fixed in certain configurations $L$ and $R$ (indicated in blue). [Top right] Corresponding quantum circuit ${\cal C}$, acting on a linear chain of quantum systems. The $q^4$ Boltzmann weights $w^{ij}_{kl}$ at each site constitute the matrix elements of a quantum gate acting on two neighboring systems (represented by a box). The boundary conditions $L$ and $R$ are mapped to basis states $|L\rangle$ and $|R\rangle$. [Bottom left and right] Vertex model on triangular lattice and corresponding quantum circuit built of 3-local gates.} \end{figure}

{\bf 2. Quantum circuits as vertex models.---} Next we show how a general class of classical spin systems, called ``vertex models'', is naturally associated with quantum circuits. A vertex model (VM) consists of classical $q$-state spins $s_e \in \{0,1,\ldots ,q-1\}$ placed on a lattice, where a spin $s_e$ is associated with each edge $e$ of the lattice, and where (many-body) interactions occur in the vertices of the lattice. The energy of a spin configuration $\mathbf{s} = (s_e)$ is given by a sum of local contributions: $H({\bm s})= \sum_a h^{a}$. Here $h^{a}$ is the interaction at site $a$, which is a function of those spins situated at all edges incident with $a$. For example, in the case of a VM on a 2D square lattice, $h^{a}=h^{a}(s_u,s_d,s_l,s_r)$ is a 4-body interaction depending on the spins ``up'', ``down'', ``left'' and ``right'' of vertex $a$ (see Fig. \ref{Fig:VM1}).

Instead of defining an energy function $H$, a VM is frequently expressed directly in terms of its Boltzmann weights $W({\bm s})=e^{-\beta H({\bm s})}$ (where $\beta$ denotes the inverse temperature).  This typically amounts to prescribing which spin configurations are allowed to have a nonzero weight. Note that $W(\bm s) = \prod_a w^{a}({\bm s})$ is a product of local weights $w^{a}(s)=e^{-\beta h^{a}(s)}$. For example, the ``eight-vertex model'' is a model on a 2D square lattice with two-state spins, where only eight out of the $2^4 = 16$ spin configurations around each vertex are allowed to have a nonzero weight. 

A central quantity for any spin system is the partition function \be {\cal Z}=\sum_{\bm s} e^{-\beta H({\bm s})},\ee since all thermodynamical properties of the system can be derived from ${\cal Z}$.

\hspace{1cm}

We now show how vertex models can be mapped to the quantum circuit model. We will illustrate the mapping for a $q$-state vertex model on a 2D tilted square lattice (see Fig. \ref{Fig:VM1}). The construction is, however, not limited to this case, as we will see below.
We choose open boundary conditions and furthermore fix the spins at the left and right boundary of the lattice (where one has, say, $N$ spins on each side) in some arbitrary configurations denoted by $L = (s_1^L, \dots, s^L_N)$ and $R=(s^R_1, \dots, s^R_N)$. In the horizontal direction, the lattice contains $M=$ poly($N$) sites. 
Note that, as every vertex of the lattice has degree 4, there are $q^4$ possible spin configurations at each site, and therefore $q^4$ possible Boltzmann weights $w^{a}(i,j, k,l)$; here $i, j, k, l$ denote the four spins at the site $a$. To simplify notation we will denote the Boltzmann weights by $w^{ij}_{kl}$, hence omitting the dependence on $a$ (but keeping in mind that Boltzmann weights are generally site-dependent). We will organize these Boltzmann weights in a $q^2\times q^2$ matrix denoted also by $w^{a}$, by grouping the spins $i,j$  ($k, l$) to be the row (column) index of the matrix: $w^a = \sum w^{ij}_{kl}|ij\rangle\langle kl|$. For example, in the case of the 8VM \cite{Ba89}, $w^{a}$ is a $4\times 4$ matrix of the form \be \label{8VM} w^{a} = \left(\begin{array}{cccc} w^{00}_{00} & 0 & 0&w^{00}_{11}\\ 0 &  w^{01}_{01} & w^{01}_{10} & 0\\ 0 &  w^{10}_{01} & w^{10}_{10} & 0\\w^{11}_{00} & 0 & 0&w^{11}_{11}\end{array} \right). \ee
The six-vertex model \cite{Ba89} is obtained by additionally setting $w^{00}_{11} =0 =w^{11}_{00}$.

The connection to the quantum circuit model is immediately obtained by considering the matrices $w^{a}$ as 2-local (possibly non-unitary) quantum gates acting on pairs of nearest-neighbor $q$-level quantum systems. More particularly, we  consider a linear chain of $N$ $q$-level systems, and a quantum circuit ${\cal C} = \prod_a w^{a}$ acting on this system as depicted in Fig. \ref{Fig:VM1}.
It is then an easy exercise to show the following identity: \be\label{partition} {\cal Z}_{\mbox{\scriptsize{vm}}}^{L, R} = \langle L|{\cal C}|R\rangle.\ee Here ${\cal Z}_{\mbox{\scriptsize{vm}}}^{L, R}$ is the partition function of the VM with boundary conditions $L$ and $R$, and $|L\rangle$ and $|R\rangle$ are $N$-party computational basis states associated with the spin configurations $L$ and $R$ in the obvious way.

Equation (\ref{partition}) establishes in a simple way a direct relation between VMs and the quantum circuit model. The mapping is akin to well known mappings from 2D classical to (1+1)D quantum systems; the fact that we are considering a \emph{tilted} 2D lattice, allows one to read off the quantum circuit ${\cal C}$ immediately. A number of conclusions can immediately be drawn from this correspondence. This is shown in the next two sections. 



\

{\bf  2.1. Simulable gate sets from solvable VMs.---} First, it is clear from Eq. (\ref{partition}) that every VM for which the partition function ${\cal Z}_{\mbox{\scriptsize{vm}}}^{L, R}$ can be evaluated (or approximated) efficiently, gives rise to a quantum circuit for which the matrix element $ \langle L|{\cal C}|R\rangle$ can be evaluated efficiently classically. We will illustrate the power of this mapping by giving three non-trivial examples of simulable gate sets which can be obtained in this way. The first example is obtained by considering the 8VM. It is known that the partition function ${\cal Z}_{\mbox{\scriptsize{vm}}}^{L, R}$  of this model, with possibly complex Boltzmann weights, can be evaluated efficiently for every $L, R$ and for every system size $N$, as long as the Boltzmann weights (\ref{8VM})  additionally satisfy the condition \be \label{free-fermion}  w^{00}_{00}w^{11}_{11} - w^{00}_{11} w^{11}_{00}  = w^{01}_{01}w^{10}_{10} - w^{01}_{10} w^{10}_{01}. \ee Under this condition, the model can be mapped to a system of non-interacting Fermions \cite{Fa70}, which can be solved analytically. The correspondence (\ref{partition}) immediately implies that all circuits ${\cal C}$ built out of gates of the form (\ref{8VM}), with the additional condition (\ref{free-fermion}), can be simulated efficiently classically in the sense that $ \langle L|{\cal C}|R\rangle$ can be evaluated efficiently for all basis states $|L\rangle$ and $|R\rangle$. Interestingly, this recovers a well-known result \cite{Va02}, where it was also showed that the above gates (which are called ``match-gates'') are efficiently simulable classically. The present considerations show that the simulability of match-gates is implicit in the solvability of the 8VM model, a result which has been known since long. Also the connection between match-gates and free-fermion systems, which  was (re)discovered in the context of quantum computation in \cite{fermions}, has been established for some time in the context of the 8VM (see also \cite{Hu66}).

As a second example, we consider a VM on a 2D tilted triangular lattice as in Fig. \ref{Fig:VM1}, with boundary conditions $L$ and $R$. Going through a similar argument as above, one finds that the partition function ${\cal Z}_{\mbox{\scriptsize{vm}}}^{L, R}$ of this model can again be written in terms of a quantum circuit ${\cal C}=\prod w^{a}$ as in (\ref{partition}), but where now the gates $w^{a}$ act on triples rather than pairs of $q$-level systems. Indeed, since every site in the lattice has degree 6, the Boltzmann weights are organized in a $q^3\times q^3$ matrix, such that each elementary quantum gate $w^{a}$ is a 3-local gate. The structure of the circuit is also given in Fig. \ref{Fig:VM1}. We now recall that a particular $q=2$ VM of this form, the ``32-vertex model'' is known to be solvable \cite{Sa75} (also by using a mapping to free fermions). Because of space limitations we omit the specific form of the 32VM Boltzmann weights, and we refer to e.g. \cite{Sa75}.

As a third example, we remark that there exist $q=2$ solvable (free-fermion) VMs on a 3D square lattice, see e.g. \cite{3Dvertex}. A 3D model on a tilted lattice would correspond to a simulable quantum circuit acting on a 2D array of qubits. Note that, as the degree of the 3D lattice is 6, the elementary gates in the circuit are 3-qubit gates.




\

{\bf  2.2. Quantum algorithms for VMs.---} We now consider a second application of identity (\ref{partition}), namely the construction of quantum algorithms to approximate partition functions of VMs. Also here the argument is particularly simple. The key point is that, given any poly-sized unitary quantum circuit ${\cal C}$, the quantity $ \langle L|{\cal C}|R\rangle$ can be efficiently approximated on a quantum computer using the ``Hadamard test'' (see e.g. \cite{Ah06}). Using (\ref{partition}), this immediately yields efficient quantum algorithms to approximate partition functions of VMs in those parameter regimes for which the matrices $w^{a}$ are unitary. The type of approximation which can be achieved by the Hadamard test, is an ``additive approximation''. This entails that this quantum algorithm returns as a result a number $c$ in poly$(N)$ time which satisfies \be\label{additive}
|c -  \langle L|{\cal C}|R\rangle|\leq p(N)^{-1},
\ee
with a success probability that is exponentially (in $N$) close to 1, where $p(N)$ is some polynomial. We note that  this type of approximation is achieved in many recent quantum algorithms \cite{Ah06, Ah07, Ar08}. As an example of the present technique, an efficient quantum algorithm to approximate the partition function of the 8VM (and, a fortiori, the 6VM) is obtained for all Boltzmann weights for which the matrices (\ref{8VM}) are unitary.  Note that no restriction is required regarding translation invariance of the model: the $w^{a}$'s may be different at each site, as long as these matrices are unitary. On the other hand, it is clear that, in order for $w^{a}$ to be unitary, some of the Boltzmann weights  have to be negative (or even complex) numbers. Therefore, it seems difficult to obtain with this technique approximations of partition functions with positive and thus ``physical'' Boltzmann weights. This problem seems to be present also in other work, see e.g. \cite{Ah07, Va08}, and it is at present unclear whether there exist a general method to overcome it.

We now show that our quantum algorithms are indeed non-trivial, by considering the {\em computational complexity} of the problems that can be solved with these algorithms. In particular, we will show that approximating the partition function of the 6VM in the abovementioned unitary parameter regime is as difficult as simulating \emph{all} quantum computations, and hence unlikely to be possible in an efficient way classically. More precisely, we will show that approximating the partition function ${\cal Z}^{L, R}_{\mbox{\scriptsize{6vm}}}$ of the 6VM on a (tilted) 2D square lattice, with staggered left and right boundary conditions $L = R = (0101\dots)$, and where the matrices $w^{a}$ are unitary, is a BQP-complete problem \cite{noteBQP}.
To achieve this, we need to show that every quantum computation can be reduced, with polynomial computational effort, to the approximation of ${\cal Z}^{L, R}_{\mbox{\scriptsize{6vm}}}$ as in (\ref{additive}). To prove this, we use that the exchange interaction \be H_{\mbox{\scriptsize{ex}}} =\sigma_x\otimes \sigma_x + \sigma_y\otimes \sigma_y + \sigma_z\otimes \sigma_z\ee is (encoded) universal for quantum computation \cite{exchange}. The corresponding unitary gates $U=e^{it H_{\mbox{\scriptsize{ex}}}}$ are of the form (\ref{8VM}) with non-zero entries \be w^{00}_{00}&=&w^{11}_{11}=e^{i2t}\nonumber\\ w^{01}_{01}&=&w^{10}_{10}=\cos(2t)\nonumber\\ w^{01}_{10}&=&w^{10}_{01}=i\sin(2t).\ee The required encoding uses four-qubit logical states of the form $|0_L\rangle\propto (|01\rangle -|10\rangle)^{\otimes 2}$. From this encoded universality, it can be shown (see e.g. \cite{Ah06}) that the problem of providing an additive approximation (as in (\ref{additive})) of the number $\langle0_L|^{\otimes N} {\cal C}|0_L\rangle^{\otimes N}$ is BQP-complete, where  ${\cal C}$ is an arbitrary poly-sized quantum circuit acting on $N$ logical qubits, and where each elementary gate in ${\cal C}$ has the form $e^{itH_{\mbox{\scriptsize{ex}}}}$. In order to connect this problem to the 6VM partition function, we first note that the unitary two-qubit gates $e^{itH_{\mbox{\scriptsize{ex}}}}$ in ${\cal C}$ correspond to Boltzmann weights $w^{a}$ of a 6VM. Second, we need to be careful with the boundary conditions: since $|0_L\rangle$ is an entangled state, we cannot simply interpret it as a boundary condition as in (\ref{partition}), since $|L\rangle$ and $|R\rangle$ need to be computational basis states. However, we note that $|01\rangle - |10\rangle$ can be prepared from a basis state $|01\rangle$ with a unitary gate $V$ which ``obeys'' the 6VM structure: simply take $V$ to have the structure (\ref{8VM}) with non-zero elements \be w^{00}_{00}&=&w^{11}_{11}=1,\nonumber\\ w^{01}_{01}&=&w^{10}_{10}=w^{01}_{10}=-w^{10}_{01}=1/\sqrt{2}.\ee This allows one to write \be \langle0_L|^{\otimes N} {\cal C}|0_L\rangle^{\otimes N} = \langle 01|^{\otimes 2N} {\cal C}' |01\rangle^{\otimes 2N},\ee where ${\cal C}' = [V^{\dagger}]^{\otimes 2N}{\cal C} V^{\otimes 2N}$. We can now identify $\langle 01|^{\otimes 2N} {\cal C}' |01\rangle^{\otimes 2N}$ with the partition function of a 6VM with staggered left and right boundary conditions. This proves the BQP-completeness of the 6VM.

\

\begin{figure} {\includegraphics[width=9cm]{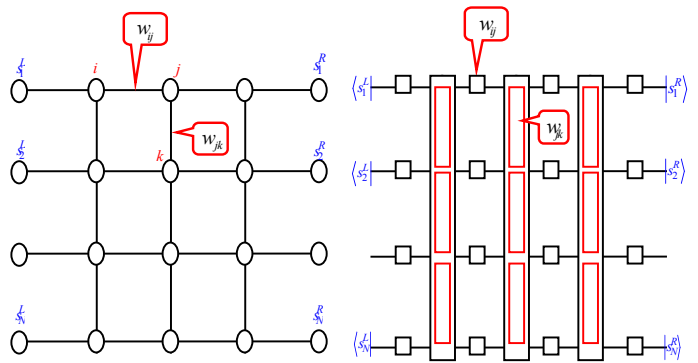}} \caption[]{\label{Fig:VM2} [Left] Edge model on a 2D square lattice. On every vertex, a $q$-state classical spin (illustrated in red) is situated. Along every edge, an interaction takes place between the two spins  $i$ and $j$ at the incident vertices, with corresponding Boltzmann weight $w_{ij}$. The spins on the left and right boundary vertices are fixed in certain configurations $L$ and $R$ (indicated in blue). [Right] Corresponding quantum circuit ${\cal C}$, acting on a linear chain of quantum systems. The $q^2$ Boltzmann weights $w_{ij}$ at each horizontal edge constitute the matrix elements of a quantum gate acting on one $q$-level system; the $q^2$ weights at each vertical edge constitute the diagonal entries of a diagonal two-local gate acting on two neighboring systems. The boundary conditions $L$ and $R$ are mapped to basis states $|L\rangle$ and $|R\rangle$.} \end{figure}

{\bf 3. Edge models and quantum circuits.---}  In analogy with the above results for VMs, next we present mappings between ``edge models'' and quantum circuits. In contrast with a VM, an edge model (EM) is a spin system on a lattice where $q$-state spins $s_a$ reside on the sites $a$ of the lattice, and two-body interactions $h^{e}(s_a, s_b)$ occur along the edges $e=ab$ of the lattice, between pairs of spins. Important examples of EMs are the Ising and Potts models \cite{Potts}. 
We will illustrate our results for an Ising-type model with 2-state spins arranged on a 2D rectangular lattice (see Fig. \ref{Fig:VM2}), but the results will be easily generalized. The energy function of such a model is $H(s)=\sum_{e} h^{e}(s_a, s_b)$ (where the sum is over all edges $e=ab$) and we denote the corresponding Boltzmann weights by $w^{e}_{ij}=\exp[{-\beta h^{e}(i, j)}]$ (with $i, j = 0, 1$). Notice that, by allowing complex coupling strengths, the Boltmann weights $w^{e}_{ij}$ can take arbitrary (complex) values. The dimensions of the lattice are $N\times M$, where $M =$ poly$(N)$ is the horizontal dimension, as before, but this time we do \emph{not} consider  a tilted lattice.   Finally, we fix left and right boundary conditions $L$ and $R$.

We now associate with this spin system a quantum circuit in the following way. Consider a linear chain of $N$ qubits and a quantum circuit ${\cal C}$ consisting of $M$ ``layers'' of one- and two-qubit gates, which has a structure as depicted in Fig. \ref{Fig:VM2}. There are two types of layers in ${\cal C}$, which alternate each other. The first type of layer is associated with a slice of horizontal edges in the lattice; it consists of $N$ single-qubit gates acting on the $N$ qubit wires, where each gate has the form $\sum_{i,j=0}^1 w_{ij}^e|i\rangle\langle j|$. That is, the four possible Boltzmann weights along an edge form the entries of a single-qubit gate.  The second type of layers is associated with a vertical slice of edges, and consists of $N-1$ nearest-neighbor, diagonal (and thus mutually commuting) two-qubit gates of the form $\sum_{i,j=0}^1 w_{ij}^e|ij\rangle\langle ij|$. Thus, here the Boltzmann weights are the diagonal elements of these diagonal gates. With this circuit ${\cal C}$, one finds that  \be\label{partition2}{\cal Z}_{\mbox{\scriptsize{em}}}^{L, R} = \langle L|{\cal C}|R\rangle.\ee Here ${\cal Z}_{\mbox{\scriptsize{em}}}^{L, R}$ is the partition function of the EM with boundary conditions $L$ and $R$, and $|L\rangle$ and $|R\rangle$ are defined as before. Thus, by suitably identifying the Boltzmann weights of a model with single- and two-qubit unitary operations,  the partition function of any EM is associated with a quantum circuit.


\hspace{1cm}


\

{\bf  3.1. Quantum algorithms for edge models.---} As in the case of VMs, Eq. (\ref{partition2}) now immediately leads to a simple quantum algorithm, via the Hadamard test, to approximate  ${\cal Z}_{\mbox{\scriptsize{em}}}^{L, R}$ in those parameter regimes for which the single-qubit gates $\sum w_{ij}^e|i\rangle\langle j|$ and two-qubit gates $\sum w_{ij}^e|ij\rangle\langle ij|$ are unitary. Note that a unitary two-qubit gate is obtained if and only if $|w^{e}_{ij}|=1$.

Moreover, it is again possible to prove that these algorithms are associated with BQP-complete problems. To show this, one remarks that universal quantum computation can be achieved by quantum circuits of the form considered here. In fact, universality is already obtained when considering a suitable discrete set of unitary gates---corresponding to a discrete set of (complex) couplings in the corresponding EM. Indeed, it is well-known that the single-qubit gates \be I_1 &=& |0\rangle\langle0| + |1\rangle\langle1|\quad \mbox{(identity)},\nonumber\\ H&\propto& \sum(-1)^{i\cdot j}|i\rangle\langle j| \quad\mbox{(Hadamard)},\nonumber\\ P &=& |0\rangle\langle0| + e^{i\pi/8}|1\rangle\langle1| \quad\mbox{(phase gate)},\ee together with the two-qubit identity and nearest-neighbor controlled-phase gate \be I_2 &=& \sum_{i,j}|ij\rangle\langle ij|\nonumber\\ CP &=& \sum_{i,j}(-1)^{i\cdot j}|ij\rangle\langle ij|,\ee  are sufficient to achieve universal quantum computation. This implies that the approximation (in the sense of (\ref{additive})) of an Ising-type partition function  ${\cal Z}_{\mbox{\scriptsize{em}}}^{L, R}$ on a 2D square lattice, with possible Boltzmann weight associated with the above universal gate set, and with boundary conditions $L = R = (000\dots)$ (where all boundary spins are pointing in the same direction), is a BQP-complete problem.

Note that the BQP-completeness of this 2D Ising-type
model (and hence the existence of an efficient quantum algorithm to
approximate the problem) is remarkable, as it is known that the exact evaluation of the 2D Ising spin
glass in the physical parameter regime is associated with NP-complete problems \cite{Ba82}. 

We finally note that a similar quantum algorithm as above can be obtained for the partition function of the standard 2D Ising model with magnetic fields, although in this case one needs to take a number of pre-factors into account in the relation (\ref{partition2}) between the partition function and the quantum circuit.


\

{\bf  3.2. Simulable gates sets for edge models.---} By considering exactly solvable EMs, such as the 2D Ising model without magnetic field, the mapping (\ref{partition2}) allows one to directly read off simulable gate sets.
For the 2D Ising model without external field, which has an energy function satisfying $h^e(0,0) = h^e(1,1)$ and $h^e(0,1) = h^e(1,0)$, the corresponding single-qubit gates of the form $\exp({i\alpha \sigma_x})$, and the two-qubit diagonal gates of the form $\exp({i\beta \sigma_z\otimes\sigma_z})$. Invoking that the partition function of such a 2D Ising model can be efficiently evaluated (the latter can again be proved by mapping this classical system to a quantum system of non-interacting fermions), we can conclude that every circuit ${\cal C}$ built out of $\sigma_x$-rotations and nearest-neighbor ${\sigma_z\otimes \sigma_z}$-rotations, can be efficiently simulated classically (in the sense that $\langle L| {\cal C} |R\rangle$ can be evaluated efficiently).

\

{\bf  4. Summary and outlook.---} We have established simple but general mappings between quantum circuits and classical vertex and edge models. This framework has been used  (i) to derive simulable gate sets for quantum computation from exactly solvable classical models; and (ii) to derive quantum algorithms to approximate the partition function of classical models in certain parameter regimes, showing that approximating certain  partition functions is as hard as simulating universal quantum computation.  Our results are not limited to the examples given in this paper, and can easily be extended.


The investigation initiated here is by no means complete, and faces several interesting challenges. 

Regarding (i), we note that all solvable models presented here are solvable via mappings to free fermions. The major challenge is to also incorporate in our framework the powerful formalism of the Yang-Baxter equation (YBE) and the Bethe ansatz in order to study the classical simulation of quantum computation. In this context one needs to be careful to take into account that the YBE typically provides solutions for classical models in the thermodynamic limit, and in a limited regime of parameters, and often for systems with a high degree of translational symmetry. These limitations imply that the YBE cannot trivially be incorporated in the present framework, and some additional work is required; this is an ongoing investigation. In this context, it is also important to remark that in this paper we have only considered classical models for which the partition function can be evaluated exactly. However, for an efficient simulation of the corresponding quantum circuits, an approximate solution would be sufficient. Our current investigations indicate that such approximations may play a role when employing the YBE in this study.

We also mention that many of the simulability results in this paper are not restricted to unitary quantum circuits: efficient classical simulation of certain non-unitary circuits can be achieved in a similar way (describing e.g. post-selected quantum computation). 

Regarding (ii),  we believe that the main merit of the obtained quantum algorithms is twofold. First, proving  BQP-completeness of a spin model, as we have done for the 2D six-vertex model as well as a 2D Ising-type model, strongly suggests that ``solving'' such a model is a computationally difficult task. We may therefore learn something about the complexity of spin systems via the present results. Second, BQP-complete problems by definition capture what the power of quantum computation is. By relating quantum computation to well studied classical spin systems, we believe that new insights may be gained about the power of quantum computers.

The main challenge in the context of these quantum algorithms is clear: it would be of great interest to obtain quantum algorithms which can treat classical spin systems in their ``physical'' parameter regime, i.e., considering real and positive Boltzmann weights. This would make it possible to use a quantum computer to simulate classical spin systems. We are currently studying this issue within the framework presented in this paper.











{\em Acknowledgements:} This work was supported by: the European Union (QICS, OLAQUI, SCALA) and the FWF (W.D and H.J.B.), the excellence cluster MAP (M.V.D.N.), and NSERC (R.R.).

After completing this work, we became aware of other recent work regarding quantum algorithms to approximate partition functions of classical spin systems \cite{Ar08}; their investigation is related to our work \cite{Va08}.






\end{document}